\documentclass[12pt]{article}
\usepackage{graphicx}
\usepackage{natbib}
\bibpunct{(}{)}{,}{a}{}{,}
\usepackage{geometry} 
\title{Student's {\it t\/}-Distribution Based Option Sensitivities:\\
Greeks for the Gosset Formulae}

\author{Daniel T. Cassidy$^{\dag ,}$\thanks{Corresponding author. Email: cassidy@mcmaster.ca}, 
Michael J. Hamp $^\ddag$, and 
Rachid Ouyed $^\S$ \\
 $^\dag$Department of Engineering Physics, McMaster University,\\
Hamilton, ON, L8S 4L7, Canada\\ 
cassidy@mcmaster.ca\\
$^\ddag$ Scotiabank, Toronto, ON, M5H 1H1, Canada\\
mike\_hamp@scotiacapital.com \\
$^\S$ Department of  Physics \& Astronomy, University of Calgary,\\Calgary, AB, T2N 1N4,
Canada, and   \\
Origins Institute, McMaster University,\\Hamilton, ON, L8S 4M1 Canada\\
rouyed@ucalgary.ca
}

\begin{document}

\maketitle

\begin{abstract}
European options can be priced when returns follow a Student's
 {\it t\/}-distribution, provided that the asset is capped in value or the distribution is truncated.  We call pricing of options using a log Student's {\it t\/}-distribution a Gosset approach, in honour of W.S. Gosset.  In this paper, we compare the greeks for Gosset and Black-Scholes formulae and we discuss implementation.  The {\it t\/}-distribution requires a shape parameter $\nu$ to match the ``fat tails'' of the observed returns.  For large $\nu$, the Gosset and Black-Scholes formulae are equivalent.  The Gosset formulae removes the requirement that the volatility be known, and in this sense can be viewed as an extension of the Black-Scholes formula. 
\end{abstract}
\indent
\hspace*{1.28cm}{\it Keywords}: option pricing; Student's $t$; fat tails.

\section{Introduction}
\hspace*{1.28cm}It is known that Student's {\it t\/}-distributions fit the returns of stocks and equity
indices better than normal distributions \citep{man, fam, pra, bla, dej, pla, zhu, ger, huw}. Prices for European options can be calculated for returns that follow a Student's {\it t\/}-distribution if the value of the underlying asset is capped or if the {\it t\/}-distribution is truncated \citep{bou, mcc, cas}.  Capping the value of the asset or truncating the distribution keeps the integrals, which are required to price the option, finite.

\hspace*{1.28cm}Truncation is not an academic exercise.  It is unphysical for the value of an
asset to approach infinity \citep{bou}.  Thus truncation is required for a model of values of
assets and of returns on assets to reflect reality.  Capping has a similar effect as truncation and is simply a slightly different approach to creating a physical model for the price of an option.

\hspace*{1.28cm}It is the ``fat tails'' of the Student's {\it t\/}-distribution that make truncation
necessary \citep{bou, mcc, cas}.  The normal pdf, on which the Black-Scholes formula for the price of an
European option is based, decreases rapidly for values that are more than several standard deviations beyond the mean value.  This rapid decrease keeps the integrals, which are required to price an option, finite and thus allows a price to be found without consideration of truncation or capping.  However, this rapid decrease in the tails is not consistent with the observed distribution of returns, and thus calculations based on a normal pdf underestimate the probability of events that are several standard deviations beyond the mean.

\hspace*{1.28cm}The Student's {\it t\/}-distribution offers support from $-\infty$ to +$\infty$ \citep{eva}.  However there is no requirement that the full domain be used when working with a distribution.  If the data do not or can not exist over the full domain and yet the frequency of occurrence of the data do follow the distribution over a subset of the full domain, it is correct to truncate the domain to match the observations or physical constraints.  One need only use a concept from conditional probability to maintain the normalization of the truncated pdf.  In terms of the stock market, the magnitude of the returns and the value of an asset are truncated or capped.  In practice, one does not observe a stock that is worth an infinite amount or daily returns on indices that are $>$ 30\%.

\hspace*{1.28cm}In working with a Student's {\it t\/}-distribution to price European options, it is
necessary to decide upon the shape parameter $\nu$ and the scale parameter $\sigma_{{\rm {\it T\/}}}$ (the scale parameter $\sigma_{{\rm {\it T\/}}}$ is the volatility) for the {\it t\/}-distribution, and to decide where to place the cap or the truncation.  The decision on the value for the cap or truncation
is based to a large extent on tolerance to risk or equivalently on the anticipated maximum value for the asset.  One can choose to work with, e.g.,  $p = 99.999$\% confidence or 95\% confidence, or choose the corresponding maximum value {\it x$_{{\rm c}}(p)$\/}.  The shape parameter $\nu$ and the volatility $\sigma_{{\rm {\it t\/}}}$ are a different story.  These parameters should be chosen to match the underlying distribution.

\hspace*{1.28cm}In this paper we discuss pricing options when the underlying distribution is a
Student's {\it t\/}-distribution.  We call a model to price European options when the underlying distribution is a {\it t\/}-distribution a Gosset model, in honour of W.S. Gosset who published under the name Student \citep{zab}.  We briefly discuss implementation, present the Gosset equations in Appendix B, discuss what is an appropriate shape parameter $\nu$ based on historical returns, and compare some of the greeks for the Gosset and Black-Scholes prices for European call options.

\bigskip
\section{Implementation}

\hspace*{1.28cm}The Gosset formulae for the prices of European options are similar to the
Black-Scholes formula.  In the limit as the shape parameter $\nu$ approaches infinity, the Gosset formulae become the Black-Scholes formula.  Both formulae are based on tabulated functions.  The Black-Scholes formula requires values for the error function.  The Gosset formulae require tabulated values for integrals of log Student's {\it t\/}-distributions.  Fortunately both the error function and the values for integrals can be calculated numerically in short time periods.  

\hspace*{1.28cm}Appendix A is a copy of the output of a Maple worksheet that was used to
determine the time required to calculate and print to the screen the value of a call option for 100 different values for the asset {\it S\/}.  This script took 20 s to execute.  One hundred evaluations of the Maple intrinsic function blackscholes({\it S\/}, {\it K\/}, {\it r\/}, {\it T\/}, $\sigma$) for 100 different values of {\it S\/} would take 5 ms to complete and to print to the screen.  The execution time for pricing a call with a Gosset formula could be reduced if optimized code is used to evaluate the integrals, similar to the optimized code that is used for evaluation of the error function to price an option using the Black-Scholes formula.

\hspace*{1.28cm}One challenge with use of the Gosset formula is that spreadsheets do not have built-in functions to evaluate the integrals, unlike the functions that can be used to price an option with the Black-Scholes formula.  However, in time this shortcoming can be remedied.

\hspace*{1.28cm}One advantage that the Gosset formulae enjoy over the Black-Scholes formula is the inclusion of the ``fat tails'' that returns display.  The Gosset formulae can be viewed as extensions to the Black-Scholes formula.  The Gosset formulae price European options where the returns are normally distributed but the true (parent) volatility is not known {\it a priori \/}(i.e., the true volatility is estimated as the sample standard deviation) or the true volatility is a random variable and is distributed as an inverse chi variate.  The Black-Scholes formula is based on a volatility that is known; there is no uncertainty in the volatility in the Black-Scholes approach. It is known that it is important to include stochastic volatility in pricing options -- see, {\it e.g.}, \citep{hes, bak, mor, huw}.   

\hspace*{1.28cm}For the Gosset formulae to hold strictly, the volatility must be distributed as
a chi random variable or as an inverse chi variate.  The requirement of a chi or inverse chi pdf depends on the approach that is chosen to justify theoretically the fact that returns follow a Student's {\it t\/}-distribution \citep{pra, ger}.  One can postulate that the Student's {\it t\/}-distribution and the concomitant fat tails arise from the uncertainty introduced by estimating the true volatility from the sample standard deviation, or one can postulate that the volatility is a random variable.  Presumably the assumption that the volatility is a random variable takes into account the uncertainty in estimating the true volatility from the sample standard deviation.  

\hspace*{1.28cm}Historical data show that the volatility follows, to a good approximation, an inverse chi distribution \citep{pra, ger}, which suggests that the true volatility is a random variable.  The volatility is not constant in time.  The markets show periods of calm and of wild swings in prices.  These periods of calm and wild swings in price can not be predicted with certainty and thus the volatility is a random variable to an observer of the market.  

\hspace*{1.28cm}We find that both the inverse chi and the chi distributions fit the volatility
data for the S\&P 500 and the DJIA, with the inverse chi distribution providing a
moderately better description of the volatility than the chi distribution.  Thus one
would expect the returns to be described by a Student's {\it t\/}-distribution and not a
normal pdf, as is observed.  Hence the predictions of the Gosset formulae should
be preferred over the predictions of the Black-Scholes formulae for the prices of
European options.

\hspace*{1.28cm}Figure 1 gives the price of call and put options as a function of the strike
price {\it K\/}, asset value {\it S\/}$_{{\rm 0}}$ = 50.00, confidence level {\it p\/} = 0.999, risk free rate $r$, time to expiration $T$, {\it r\/}$\times${\it T\/} = 0.03, volatility $\sigma_{{\rm {\it T\/}}}$ = 0.3, and various values of the shape parameter $\nu$ for the Student's {\it t\/}-distribution.  The prices for $\nu = \infty$ were calculated with the Black-Scholes formula.  The solid curves for $\nu$ = 3, 8, and 21 were calculated using the Gosset formulae, Eqs. (B10) and (B11) of Appendix B, for a truncated pdf.  The broken lines were calculated using the Gosset formulae, Eqs. (B6) and (B7) of Appendix B, for a capped distribution.  Clearly the difference between the Gosset price and the Black-Scholes price increases as $\nu$ becomes smaller.  As $\nu$ becomes smaller, the tails of a Student's {\it t\/}-distribution become ``fatter''.  It is also clear from Fig. 1 that the capped Gosset price, i.e., the price for an option that is calculated using a value for an asset that is capped and follows a log Student's {\it t\/}-distribution, is marginally greater than the truncated Gosset price.  The additional cost for a capped value allows for the possibility that the value of the asset may exceed the cap value.  Truncation assumes that the value of the asset will not exceed the cap.

\begin{figure}
\begin{center}
\label{fig:fig1}
\includegraphics[scale=0.5]{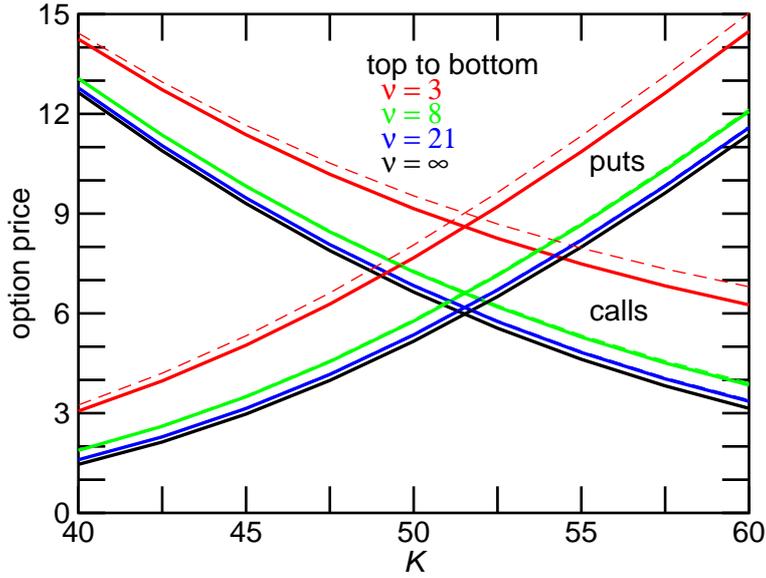}
\caption{Plot of the option price as a function of the strike {\it K\/}.}
\end{center}
\end{figure}

\hspace*{1.28cm}Table 1 gives the critical values $x_c(p)$ and increase in the value of the asset at the critical value for {\it p\/} = 0.999 for four different values of the shape parameter $\nu$.  The critical value $x_c$ and confidence level $p$ are defined such that the probability of drawing a return $\leq x_c$ is $p$.  

To price an option with either the Black-Scholes approach or the Gosset approach, the value of the asset is assumed to vary as exp($\sigma_{{\rm{\it t \/}}}\,\xi$), i.e., the logarithm of the value of the asset is distributed as $\xi$.  Table 1 shows that truncating the Student's {\it t\/}-distribution at {\it p\/} = 0.999 with $\sigma_{{\rm{\it T\/}}}$ = 0.3 allows for a maximum increase of the value of the asset of 21.421 times for $\nu$ = 3.  For the normal distribution $\nu$ = $\infty$ and at {\it x$_{{\rm c}}$\/} the relative increase in the value of the asset is 2.527 times.  Table 1 also shows that the truncation need not be restrictive, and thus pricing of options with the Gosset formula is realistic.  It is worth noting that one could truncate or cap at different values of {\it p\/}, say {\it p\/} = 0.99999, which would allow for a maximum increase of the value of the asset of 106 times for $\nu$ = 3, $\sigma_{{\rm {\it T\/}}}$ = 0.3, or say {\it p\/} = 0.95, which would allow for a maximum increase of the value of the asset of 2.075 times for $\nu$ = 3, $\sigma_{{\rm {\it T\/}}}$ = 0.3 \, The choice of the level for the cap or the truncation might depend on the underlying asset.  Penny stocks might increase dramatically in terms of percent increase, but little in terms of absolute value, and thus require a large critical value to capture the possible changes.

\begin{table}
\begin{center}
\caption{Critical values and maximum increase in the value of an asset for {\it p\/} =
0.999 and $\sigma_{{\rm {\it T\/}}}$ = 0.3 for various shape parameters.}
{\begin{tabular}
{|c|c|c|c|c|} \hline
 &  $\nu$ = 3 & $\nu$ = 8 &  $\nu$ = 21 &  $\nu = \infty$ \\ \hline
critical value, {\it x$_{{\rm c}}$\/} & 10.215 & 4.501 & 3.527 & 3.090 \\ \hline
exp(0.3$\times${\it x$_{{\rm c}}$\/}) & 21.421 & 3.858 & 2.881 & 2.527 \\ \hline
\end{tabular}}
\end{center}
\end{table}

\bigskip


\section{ Estimation of the Shape Parameter $\nu$}

\hspace*{1.28cm}One question that arises naturally in pricing an option with a Gosset formula
is what value to use for the shape parameter $\nu$.  Under the assumption that the returns are normally distributed, $\nu <$ {\it N\/} where {\it N\/} is the number of independent samples that have been used to estimate the volatility.  The argument for the maximum value is straightforward.

\hspace*{1.28cm}Let {\it s\/} be the sample standard deviation that has been calculated for {\it N\/}
independent samples from a normally distributed parent population.  The sample
standard deviation {\it s\/} follows a chi distribution with $\nu$ = {\it N\/} $-$1 degrees of freedom,
$\chi_{{\rm \nu}{}}$({\it s\/}), for data that are normally distributed \citep{eva, kre}.  Any measurement of {\it s\/} has an uncertainty and this uncertainty in {\it s\/} is reflected in the shape parameter of the sampling distribution.  For normally distributed samples where the standard deviation of the parent distribution is estimated from the sample population, the sampling distribution is a Student's {\it t\/}-distribution with $\nu$ degrees of freedom, {\it f$_{{\rm s}}$\/}({\it x,~\/}$\nu$). 

\hspace*{1.28cm}The uncertainty in the estimation of the true (parent) standard deviation can
be significant.

\hspace*{1.28cm}The (sample) volatility is the sample standard deviation for the daily returns,
and a symmetric 95\% confidence interval for the true (parent) volatility $\sigma$ is \citep{kre}: 0.77$\times${\it s\/} $\leq\sigma <$ 1.43$\times${\it s\/} for $\nu$ = 21; 0.83$\times${\it s\/} $\leq$ $\sigma$ $<$ 1.27$\times${\it s\/} for $\nu$ = 43; and, 0.87$\times${\it s\/} $\leq$ $\sigma <$ 1.17$\times${\it s\/} for $\nu$ = 87, where {\it s\/} is the sample standard deviation
(volatility) with $\nu$ degrees of freedom that has been calculated from the daily
returns and it has been assumed that the returns are normally distributed.  

\hspace*{1.28cm}The maximum value for $\nu$, assuming that the volatility is estimated as the sample
standard deviation of the returns, is obtained when the volatility is constant in time.  Praetz \citep{pra} and Gerig {\it et al.} \citep{ger} found that the variance of returns
is a random variable that changes slowly in time.  Our simple analysis of historical
returns, which we present below, agrees with the findings of  Praetz \citep{pra} and Gerig {\it et al.} \citep{ger}.
This randomness in the volatility adds weight to the tails of the distribution of the
returns.

\hspace*{1.28cm}Additional uncertainty in the volatility leads to a lower value (i.e., fatter tails
for the underlying distribution) for the shape parameter in the Gosset formula.  The
additional uncertainty arises when one wishes to price an option for exercise at
some time {\it T\/} in the future.  Since the volatility is a random variable, the volatility at
time {\it T,\/} $\sigma_{{\rm {\it T \/}}}$, is unknown and the additional uncertainty must be taken into account.

\hspace*{1.28cm}We estimate values for the shape parameter $\nu$ from historical data from the
DJIA (Oct 1928 to Feb 2009) and S\&P 500 (Jan 1950 to Feb 2009) equity indexes.

\hspace*{1.28cm}In Tables 2 to 5 are mean and median values for 22 day volatilities.  For the
simulations, 22 samples were generated and the volatility (i.e., the sample standard
deviation of the 22 samples) was calculated.  The process was repeated 1000 times,
and the mean, median value, and the standard deviation {\it s\/} of the 1000 22 day
volatilities were calculated and are reported in the tables.

\hspace*{1.28cm}For each group of 22 values, the minimum value and the maximum value
were dropped to give 20 values, and the volatility was calculated for this truncated
set of data.  The process was repeated 1000 times and the mean, median value, and
the standard deviation of the 1000 volatilities were calculated, as for the full set of
22 values.

\hspace*{1.28cm}For each group of 20 values, the minimum value and maximum value were
dropped to give 18 values.  The calculations described above were repeated and the
numbers entered into the tables.

\hspace*{1.28cm}For calculation of the expected volatility, the pdf was assumed to be
truncated at a critical value {\it x$_{{\rm c}}$\/} such that the probability of an observation {\it x\/} in the range $-${\it x$_{{\rm c}}$\/} $\leq$ {\it x\/} $\leq$ {\it x$_{{\rm c}}$\/} was equal to {\it N\/}/22, i.e., P\{$-${\it x$_{{\rm c}}$\/} $\leq$ {\it x\/} $\leq$ {\it x$_{{\rm c}}$\/} \} = $ p_N = N/22$.  There likely is a rigorous method to calculate the standard deviation for dropping the minimum
and the maximum values from a finite number of random draws from a parent population, but the method that was adopted was simple and shows the trend. 

\hspace*{1.28cm}The expected volatility $\langle \sigma \rangle$ was calculated as 

\begin{equation}
\langle \sigma \rangle = \sqrt{\int_{-x_{c} }^{x_{c} } {\frac{\xi^{2}  f (\xi  )\, d \xi }{p_N}} } ~~~{\rm with}~~ p_N = \int_{-x_{c} }^{x_{c} } f (\xi )\, d \xi 
\end{equation}
where {\it f\/}($\xi$) is the appropriate probability density function and the denominator $p_N$
maintains the normalization of the truncated pdf.

\hspace*{1.28cm}Table 2 gives results of the simulation and calculation for the procedures
described above.  Note that the normal distribution has been scaled to have the
same variance as the {\it t\/}-distribution.  Note also that the median 22 day volatility
decreases as the maximum and minimum samples are dropped, and that the median
value for the {\it t\/}-distribution decreases faster than the median value for the normal
distribution as the maximum and minimum samples are dropped.  The standard deviation of the 1000 22-day volatilities decreases greatly for the {\it t\/}-distribution as the maximum and minimum samples are dropped, but changes little for the normal distribution.  This owes to the fat tails of the {\it t\/}-distribution with $\nu$~=~3.

\begin{table}
\begin{center}
\caption{Simulated medians and standard deviations of 22 day volatilities
for samples drawn from a {\it t\/}-distribution with $\nu$ = 3 and from a normal pdf
with variance = 3, and the expected volatility $\langle \sigma \rangle$.}
{\begin{tabular}
{|c|c|c|c|c|c|c|} \hline
 &  \multicolumn{4}{|c|}{simulation}  & \multicolumn{2}{|c|}{calculation $\langle \sigma \rangle$} \\ \hline
{\it N\/} &  \multicolumn{2}{|c|}{{\it t\/}, $\nu$ = 3}  & \multicolumn{2}{|c|}{normal, $\sigma^{{\rm 2}}$=3} &  {\it t\/}, $\nu$ = 3 & normal, $\sigma^{{\rm 2}}$=3 \\ \hline
 & median & {\it s\/} & median & {\it s\/} & $\langle \sigma \rangle$ & $\langle \sigma \rangle$ \\ \hline
22 & 1.48  & 0.65 & 1.70  & 0.28 & 1.73 & 1.73 \\ \hline
20 (drop 2) & 1.07  & 0.25 & 1.41  & 0.25 & 1.01 & 1.39 \\ \hline
18 (drop 4) & 0.87  & 0.19 & 1.21  & 0.24 & 0.816 & 1.18 \\ \hline
\end{tabular}}
\end{center}
\end{table}

\bigskip
\hspace*{1.28cm}Table 3 shows the means,  the medians, and the standard
deviations for 1000 samples for the same {\it t\/}-distribution and normal distribution in
Table 2.  The means are in agreement with the expected values in Table 2
and show the same progression as the median when the minimum and maximum
values are removed from each sample of 22 data points.

\begin{table}
\begin{center}
\caption{Simulated mean, standard deviation {\it s\/}, and median of 22 day
volatilities for samples drawn from a {\it t\/}-distribution with $\nu$ = 3 and from a
normal pdf with variance = 3.}
{\begin{tabular}
{|c|c|c|c|c|c|c|} \hline
 &  \multicolumn{3}{|c|}{{\it t\/} (simulated), $\nu$ = 3} &  \multicolumn{3}{|c|}{normal (simulated), $\sigma^{{\rm 2}}$=3} \\ \hline
{\it N\/} & mean & {\it s\/} & median & mean & {\it s\/} & median \\ \hline
22 & 1.61 & 0.65 & 1.48 & 1.71 & 0.27 & 1.70 \\ \hline
20 (drop 2) & 1.09 & 0.25 & 1.07 & 1.43 & 0.24 & 1.41 \\ \hline
18 (drop 4) & 0.88 & 0.19 & 0.87 & 1.22  & 0.23 & 1.21 \\ \hline
\end{tabular}}
\end{center}
\end{table}
\bigskip
\hspace*{1.28cm}The data in the tables show that the fat tails of the {\it t\/}-distribution result in
large uncertainties of the volatility (i.e., the sample standard deviation {\it s\/} of the 22
day volatility is large for the {\it t\/}-distribution).  The fat tails are also evident in the
difference between the median and the mean values for {\it N\/} = 22.  Note that the
uncertainty in the volatility, as measured by the standard deviation {\it s\/}, reduced
dramatically for the data from a {\it t\/}-distribution after truncation of the minimum and
maximum values, but barely changed for a normal distribution.

\hspace*{1.28cm}Table 4 compares values of the volatility for the S\&P 500 daily returns to
the simulations.  The S\&P data was broken into 676 non-overlapping segments of
22 values.  The values were sorted and then the volatility (i.e., the sample standard
deviation) was calculated for the middle 22, 20, and 18 values of the 22 returns. 
The mean, median, and standard deviation {\it s\/} of the volatilities were calculated over
the 676 volatilities.  For ease of comparison, each column was normalized by the
value for {\it N\/} = 22.

\hspace*{1.28cm}Table 5 compares means and medians of the 22 day volatilities for the
DJIA equity index to simulations.

\begin{table}
\begin{center}
\caption{Comparison of the normalized S\&P 500 mean and median 22 day
volatilities (676 values) to the normalized expected volatilities.}
{\begin{tabular}
{|c|c|c|c|c|} \hline
 & \multicolumn{2}{|c|}{S\&P 500} & \multicolumn{2}{|c|}{expected volatility $\langle \sigma \rangle$}  \\ \hline
{\it N\/} & mean & median & {\it t\/}, $\nu$ = 3 & normal \\ \hline
22 & 1 & 1 & 1 & 1 \\ \hline
20 (drop 2) & 0.803 $\pm$ 0.010 & 0.81 & 0.584 & 0.803 \\ \hline
18 (drop 4) & 0.677 $\pm$ 0.011 & 0.68 & 0.471 & 0.683 \\ \hline
\end{tabular}}
\end{center}
\end{table}
\bigskip

\begin{table}
\begin{center}
\caption{Comparison of the normalized DJIA 22 day volatilities (918 values) to
simulations.}
{\begin{tabular}
{|c|c|c|c|c|} \hline
 & \multicolumn{2}{|c|}{DJIA} & \multicolumn{2}{|c|}{simulations} \\ \hline
{\it N\/} & mean & median & {\it t\/}, $\nu$ = 3 &\shortstack{ normal (mean \\ or median) }\\ \hline
22 & 1 & 1 & mean $|$ median & 1 \\ \hline
20 (drop 2) & 0.799 $\pm$ 0.008 & 0.802 & 0.67 $|$ 0.72 & 0.83 \\ \hline
18 (drop 4) & 0.675 $\pm$ 0.009 & 0.680 & 0.55 $|$ 0.59 & 0.71 \\ \hline
\end{tabular}}
\end{center}
\end{table}
\bigskip
\bigskip
\hspace*{1.28cm}Tables 4 and 5 show that the short-term DJIA and S\&P 500 volatility
behaves more like a normal pdf than a Student's {\it t\/}-distribution.  The interpretation
is that for a short term (22 returns), the returns are approximately normally distributed.  However,
the volatility changes with time, and for longer periods of time, the uncertainty in
the volatility comes into play and the distribution becomes less normal like and
more {\it t\/} like.  This is evident from the data in Table 6, where the normalized means
and medians are reported for calculations over longer time frames of 44 and 88
returns. 

\begin{table}
\begin{center}
\caption{Normalized mean and median volatilities for the DJIA (459 and 230
values) and the S\&P 500 (338 and 169 values). } 
{\begin{tabular}
{|c|c|c|c|c|} \hline
 & \multicolumn{2}{|c|}{DJIA} & \multicolumn{2}{|c|}{S\&P} \\ \hline
{\it N\/} & mean & median & mean & median \\ \hline
44 & 1 & 1 & 1 & 1 \\ \hline
40 & 0.773 $\pm$ 0.012 & 0.792 & 0.776 $\pm$ 0.016 & 0.793 \\ \hline
36 & 0.646 $\pm$ 0.012 & 0.658 & 0.649 $\pm$ 0.016 & 0.648 \\ \hline
88 & 1 & 1 & 1 & 1 \\ \hline
80 & 0.748 $\pm$ 0.017 & 0.748 & 0.753 $\pm$ 0.023 & 0.786 \\ \hline
72 & 0.622 $\pm$ 0.016 & 0.625 & 0.626 $\pm$ 0.021 & 0.663 \\ \hline
\end{tabular}}
\end{center}
\end{table}
\bigskip
\hspace*{1.28cm}Table 6 shows that the means and medians of the truncated samples are smaller for longer
time periods.  This suggests that the data are departing from normally distributed.  

\hspace*{1.28cm}An error propagation equation was used to estimate the uncertainty in the
ratio of the means \citep{bev}.  The uncertainty in {\it B\/}/{\it A \/}was
calculated as 2$\times${\it s$_{{\rm B}}$\/}$_{{\rm /{\it A \/}}}$, where ({\it s$_{{\rm B}}$\/}$_{{\rm /{\it A\/}}}$)$^{{\rm 2}}$ = ({\it s$_{{\rm B}}$\/})$^{{\rm 2}}$/{\it A\/}$^{{\rm 2}}$ + ({\it s$_{{\rm A}}$\/})$^{{\rm 2}}$({\it B\/}/{\it A\/}$^{{\rm 2}}$)$^{{\rm 2}}$ $-$ ({\it B\/}/{\it A\/}$^{{\rm 3}}$){\it s$_{{\rm AB}}$\/}$^{{\rm 2}}$, to give a 95\% confidence level assuming that {\it A\/} and {\it B\/} are normally distributed.  The covariance {\it s$_{{\rm AB}}^{{\rm 2}}$\/} was in the range of 0.97 to 0.91 for {\it A\/} equal to the mean for the full range and {\it B\/} equal to the mean for dropping the bottom and top 5\% or 10\% of the samples used to form {\it A\/}.

\hspace*{1.28cm}Figure 2 is a plot of the normalized expected volatility as a function of the
shape parameter $\nu$ for the Student's {\it t\/}-distribution.  The long tic marks at 0.803 and
0.683 are the normalized expected volatilities for samples that are normally
distributed.  The upper curve is for  $p_N$ = 20/22 = 0.9090 while the lower curve is for
$p_N$ = 18/22 = 0.8182  The curves are normalized by the expected volatility for $p_N$ =
22/22 = 1.000

\begin{figure}[b!]
\begin{center}
\label{fig:fig2}
\includegraphics[scale=0.5]{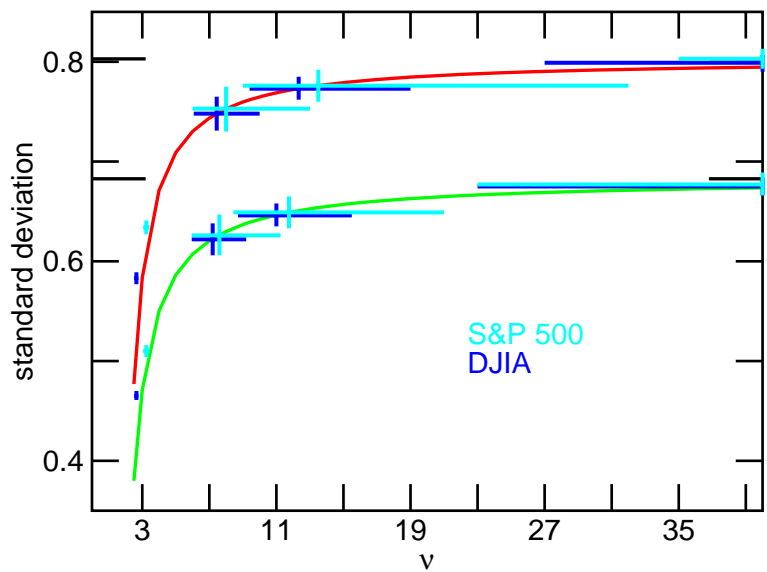}
\caption{Standard deviation of the truncated data as a function of the number of
degrees of freedom $\nu$.}
\end{center}
\end{figure}

\hspace*{1.28cm}The normalized mean volatilities for the S\&P 500 and DJIA equity indexes
were placed on Fig. 2 to match the normalized mean value of the volatility with the
normalized expected volatility curves.  This placement on the expected curve
provides a means to estimate the shape parameter and the uncertainty in the shape
parameter.  The normalized mean volatilities are found in Tables 4, 5, and 6. 
The vertical bars give the 95\% confidence intervals for the normalized mean
values.  The horizontal bars give the uncertainty in the shape parameters for the
uncertainty in the mean value.  From Fig. 2 it appears that a Student's {\it t\/}-distribution
with a shape parameter $\nu$ = 13 $\pm$ 4 would be appropriate for a 44 day volatility and
that a Student's {\it t\/}-distribution with a shape parameter $\nu$ = 8 $\pm$ 2 would be
appropriate for an 88 day volatility, based on the historical data used in the
analysis.  For a 22 day volatility, the shape parameter should be $\leq$~21.

\hspace*{1.28cm}The data on Fig. 2 at $\nu$ = 2.65 $\pm$ 0.11 and $\nu$ = 3.24 $\pm$ 0.19 were obtained from fits to the full data sets for the DJIA and S\&P 500 equity indexes \citep{cas}.  The shape parameters were obtained by fitting frequency of occurrence data to a Student's {\it t\/}-distribution.  For Fig. 2 the returns were sorted from smallest to largest and the volatility {\it s$_{{\rm A}}$\/} was calculated as the standard deviation over the {\it N\/} = 20186 returns in the DJIA data set and {\it N\/} = 14870 returns in the S\&P 500 data set.  The bottom 4.545\% and top 4.545\% of the data were
truncated, and the standard deviations {\it s$_{{\rm B}}$\/} were calculated.  Normalized standard
deviations of 0.583 $\pm$ 0.006 and 0.634 $\pm$ 0.007 were obtained by forming {\it s$_{{\rm B }}$/s$_{{\rm A}}$\/} and are plotted on the figure.  The bottom and top 4.545\% returns were truncated from the previously truncated lists, and the normalized standard deviations were
obtained.  The uncertainties in the normalized standard deviations were estimated
as 95\% symmetric confidence intervals for a chi-squared variate with degrees of
freedom equal to {\it N\/} --{} 1 \citep{kre}.  The normalized data obtained in this
manner, where the degrees of freedom were obtained from independent fits to
frequency of occurrence data, are consistent with the expected volatility curve. 

\hspace*{1.28cm}Given the above, it appears that the following model describes well the
returns, in agreement with the results of Praetz \citep{pra} and Gerig{\it et al.} \citep{ger}.  The returns are normally distributed with a defined volatility for a
short term (of order 22 consecutive returns).  The volatility is a random variable
that changes slowly in time.  A {\it t\/}-distribution, which is a mixture of a normal pdf
and a chi distribution for the reciprocal of the standard deviation (i.e., 1/volatility)
describes the pdf for the returns now and at some time in the future, where the
volatility is uncertain owing to sampling and to the slow and random development
of the volatility in time.  In mathematical notation, the sampling distribution {\it f$_{{\rm s}}$\/}({\it x, \/}$\nu$)
= $\int${\it f\/}({\it x\/}$|\sigma$) $\chi_{{\rm \nu}{}}$($\sigma^{{\rm -}{1}}$) d$\sigma$ is a Student's {\it t\/}-distribution when the parent distribution {\it f\/}({\it x\/}$|\sigma$)
is a normal distribution with standard deviation $\sigma$ and $\sigma$ follows an inverse chi
distribution (or, equivalently, the reciprocal of $\sigma$ follows a chi distribution) 
\citep{pra, ger}.

\hspace*{1.28cm}Figure 3 is a plot of a fit of a chi distribution to the reciprocal of the 22 day
volatilities for the S\&P 500 equity index from Jan 1950 to Feb 2009.  It can be
observed that the reciprocal of the historical 22 day volatilities are, for the most
part, distributed as a chi variate.  Since the returns over 22 days are found to be
normal like (see the tables) with a given volatility but the reciprocal of this
volatility is distributed as a chi variate, the distribution of the returns is expected to
follow a Student's {\it t\/}-distribution \citep{pra}.  A Student's {\it t\/}-distribution for the
returns is observed \citep{cas}.  In Fig. 3 the data deviate slightly from the best fit curve in the tail on the right hand side.  The upward sloping curves are the cumulative density functions (CDF) for the data (red) and the best fit function (black).  For a perfect fit, the black and red CDF
curves would overlap.

\begin{figure}
\begin{center}
\label{fig:fig3a}
\includegraphics[scale=0.5]{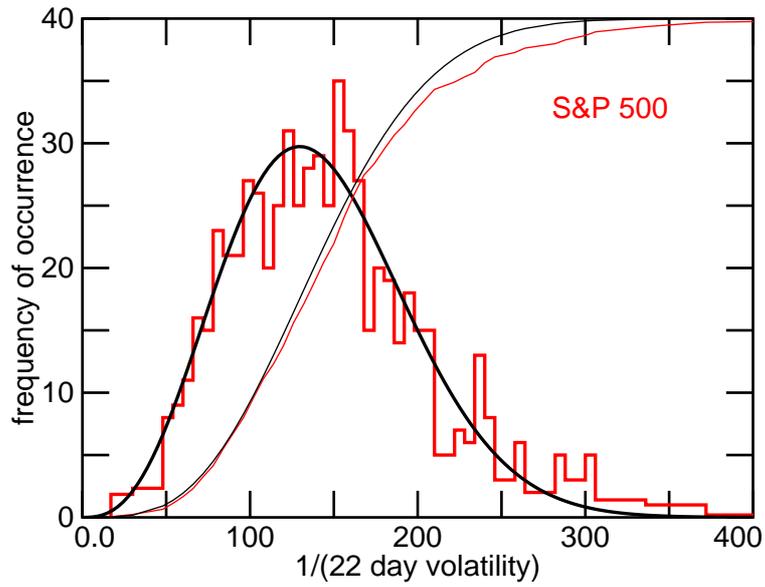}
\caption{Data and fit of a chi distribution to the frequency of occurrence of the
reciprocal of the 22 day volatilities for the S\&P 500.}
\end{center}
\end{figure}

\hspace*{1.28cm}Figure 4 is a plot of a fit of a chi distribution to the 22 day volatilities (the
volatilities, not the reciprocals of the volatilities) for the same S\&P 500 data shown
in Fig. 3.  It is interesting to note that the chi distribution fits the 22 day
volatilities reasonably well.  The fit deviates in the tails on the right hand side,
more so than the fit of Fig 3.  Fits of a chi distribution to synthetic data from an
inverse chi distribution show that the inverse chi data has a fatter tail on the right
hand side, similar to the fat tail on the right hand side that the S\&P 500 data show
in the figure.

\begin{figure}
\begin{center}
\label{fig:fig3b}
\includegraphics[scale=0.5]{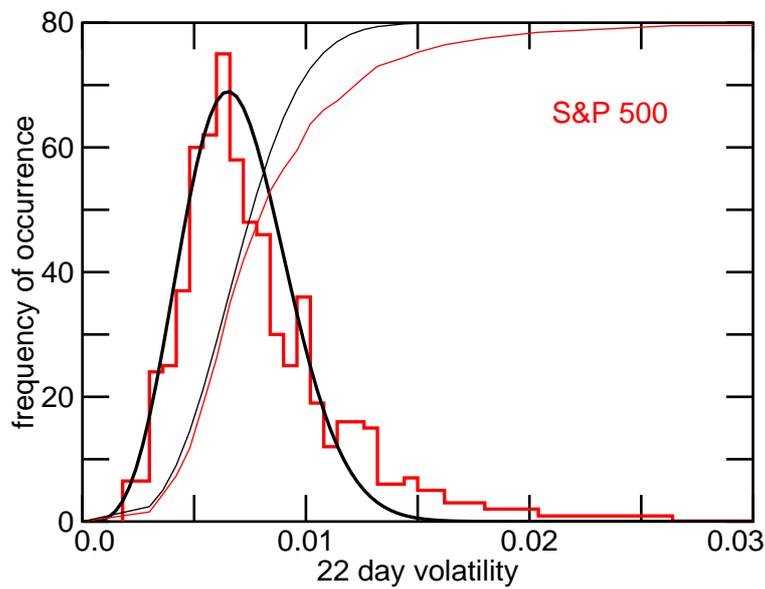}
\caption{Data and fit of a chi distribution to the frequency of occurrence of the 22
day volatilities for the S\&P 500.}
\end{center}
\end{figure}

\hspace*{1.28cm}The shape parameter for the {\it t\/}-distribution is $\nu$, the number of degrees of
freedom.  From the analysis presented here, the shape parameter is related to
estimations of the distribution of the volatility.  If the volatility does not change in
time and is known (i.e., the volatility is not estimated as the sample standard
deviations of the returns), then $\nu$ = $\infty$ and one recovers the Black-Scholes formula
for the price of an European option.  The Black-Scholes formula is derived on the
assumption that the volatility is constant and known.  If the true (parent) standard
deviation (volatility) is not known and the returns are normally distributed, then a
Student's {\it t\/}-distribution should be used.  If $\nu$ is small, then it is understood that the
true volatility might take one of a broad range of values.

\hspace*{1.28cm}The simple analysis presented also shows what might be reasonable values
(see Fig. 2) for the shape parameter of the Student's {\it t\/}-distribution.  Since the
volatility is estimated as the sample standard deviation of the daily returns, it is
expected that $\nu <$ {\it N\/}, where {\it N\/} is the size of the sample that was used to compute the
volatility.

\bigskip


\section{Greeks for Truncated Student's {\it t\/}-Distributions and Capped Values}
\hspace*{1.28cm}The Gosset formulae for the price of an option by capping the value of the
asset or by truncating the pdf are given in Appendix B.  These formulae are
extended to include a floor or a truncation at the floor level.  The notation is defined in Appendix B.

The expression for Delta for a truncated Student's {\it t\/}-distribution, $_{{\rm {\it t\/}}}\Delta$, is similar to the expression for $\Delta$ for the Black-Scholes formula.  In the limit of no truncation
({\it p$_{{\rm p}}$\/} = 0, {\it p$_{{\rm c}}$\/} = 1, {\it x$_{{\rm c}}$\/} = $\infty$) and $\nu$ = $\infty$, the expression below reduces to $\Delta$ for the Black-Scholes formula. 

\begin{equation}
{\phantom{\,} }_{t}\Delta ={\frac{ \partial  {\phantom{\,} }_{t}C_{0} }{ \partial\,\, S_{0} }} =\int_{{\frac{\ln ({\phantom{\,} }_{t} Z \times   K_{T} /S_{0}  ) - r \times T}{\sigma_{T} }} }^{x_{c} } {\frac{ e^{\, (\sigma_{T} \, \xi )}  {f}_{r}  (\xi  )\, d\xi }{{\phantom{i} }_{t} Z \times ( p_{c}-p_{p} )}}
\end{equation}
For the Black-Scholes formula, {\it $_{{\rm t }}$Z\/} = exp($\frac{1}{2}\sigma${\it $_{{\rm T}}$\/}$^{{\rm 2}}$), {\it p$_{{\rm c}}$\/} $-$ {\it p$_{{\rm p}}$\/} = 1, {\it x$_{{\rm c}}$\/} = $\infty$, and the integral can be written in terms of the error function.

The expression for $_{{\rm {\it c\/}}}\Delta$ for a capped asset price is 

\begin{equation}
{\phantom{\,} }_{c}\Delta ={\frac{ \partial  {\phantom{\,} }_{c}C_{0} }{ \partial\,\, S_{0} }} =\int_{{\frac{\ln \, ({\phantom{\,} }_{c} Z \times   K_{T} \,\, /S_{0}  ) - r \times T}{\sigma_{T} }} }^{x_{c} } {\frac{\,\, e^{\, (\sigma_{T} \, \xi  )}  {f}_{r}  (\xi  )\, d \xi }{{\phantom{i} }_{c}  Z }} +{\frac{e^{\, (\sigma_{T} \, x_{c} )} \times  (1-p_{c} )}{{\phantom{i} }_{c}  Z }}
\end{equation}

 \bigskip
\hspace*{1.28cm}Figure 5 plots $_{{\rm {\it t\/}}}\Delta$ for {\it p$_{{\rm p}}$\/} = 0, {\it p$_{{\rm c}}$\/} = 0.999, {\it K$_{{\rm T}}$\/} = \$49.00, {\it r\/}$\times${\it T\/} = 0.03, and $\sigma_{{\rm {\it T\/}}}$ = 0.3
as a function of {\it S\/}$_{{\rm 0}}$ and for $\nu$ = 3, 5, and 21.  For comparison, $\Delta$ for the Black-Scholes formula is shown.  On the scale of the figure, $_{{\rm {\it t\/}}}\Delta$ for $\nu >$ 20 lies on top of the Black-Scholes $\Delta$.  The dashed lines plot Delta for a capped distribution, $_{{\rm {\it c\/}}}\Delta$.

\begin{figure}
\begin{center}
\label{fig:fig4}
\includegraphics[scale=0.5]{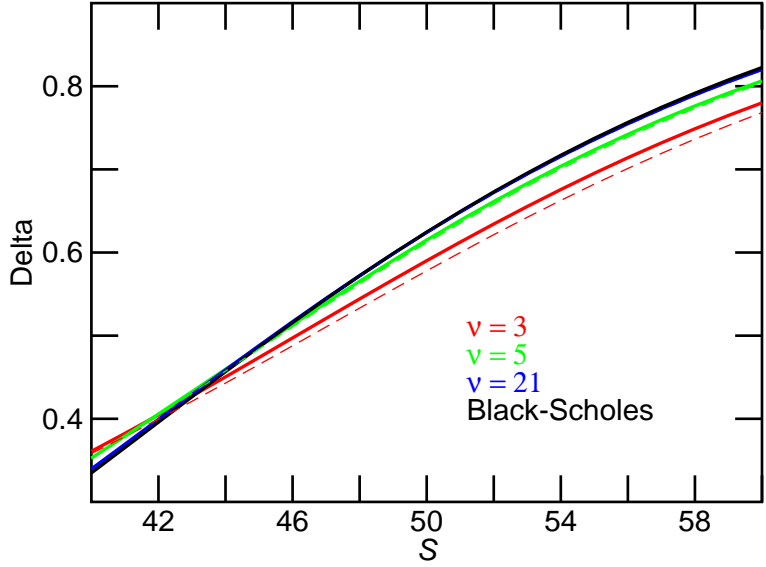}
\caption{ Delta as a function of the value of the asset $S$ for different values of the
number of the degrees of freedom $\nu$.}
\end{center}
\end{figure}

\hspace*{1.28cm}$_{{\rm {\it t\/}}}\Delta$ is a function of {\it p$_{{\rm c}}$\/}.  For large $\nu$, $_{{\rm {\it t\/}}}\Delta$ shows a weak dependence for values of
{\it p$_{{\rm c}}$\/} near unity.  This is expected since for large $\nu$ there is little difference between the Black-Scholes and Gosset formula.  The Black-Scholes formula is insensitive
to values in the tails and thus one could reasonably expect the Gosset formula for
large $\nu$ to also be insensitive to values in the tails.  Figure 6 plots $_{{\rm {\it t\/}}}\Delta$ for $\nu$ = 3{\it , p$_{{\rm p}}$\/} = 0, {\it K$_{{\rm T}}$\/} = \$49.00, {\it r\/}$\times${\it T\/} = 0.03, and $\sigma_{{\rm {\it T\/}}}$ = 0.3 as a function of {\it S\/}$_{{\rm 0}}$ and for {\it p$_{{\rm c}}$\/} = 0.99,
0.999, and 0.9999.  A small value of $\nu$ was chosen to highlight the differences. 
The dashed lines plot Delta for a capped distribution.

\begin{figure}
\begin{center}
\label{fig:fig5}
\includegraphics[scale=0.5]{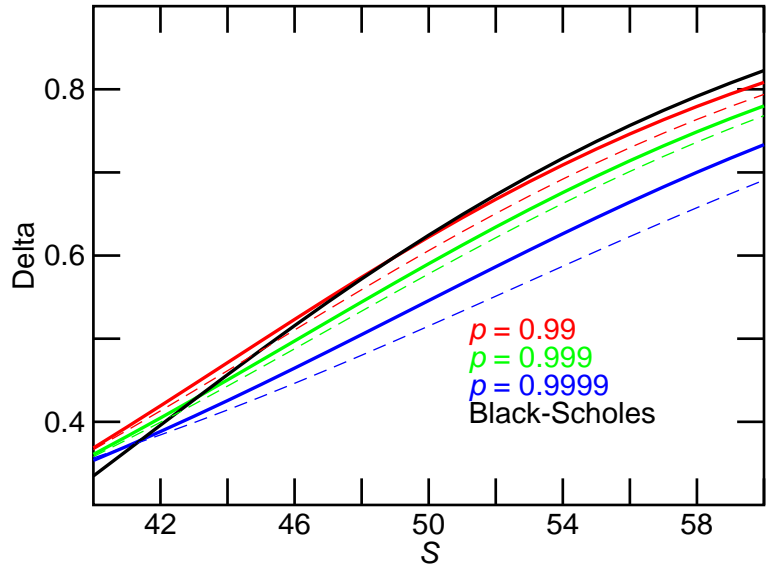}
\caption{Delta as a function of the value of the asset $S$ for different levels of
truncation $p$.}
\end{center}
\end{figure}

\hspace*{1.28cm}$_{{\rm {\it t\/}}}\Delta$ is also a function of the volatility $\sigma_{{\rm {\it T\/}}}$.  Figure 7 plots $_{{\rm {\it t\/}}}\Delta$ for $\nu$ = 3{\it , p$_{{\rm p}}$\/} = 0, {\it K$_{{\rm T}}$\/}
= \$49.00, {\it r\/}$\times${\it T\/} = 0.03, and {\it p$_{{\rm c}}$\/} = 0.999 as a function of {\it S\/}$_{{\rm 0}}$ for $\sigma_{{\rm {\it T\/}}}$ = 0.1, 0.2, and 0.3
The dashed lines give $_{{\rm {\it c\/}}}\Delta$.

\begin{figure}
\begin{center}
\label{fig:fig6}
\includegraphics[scale=0.5]{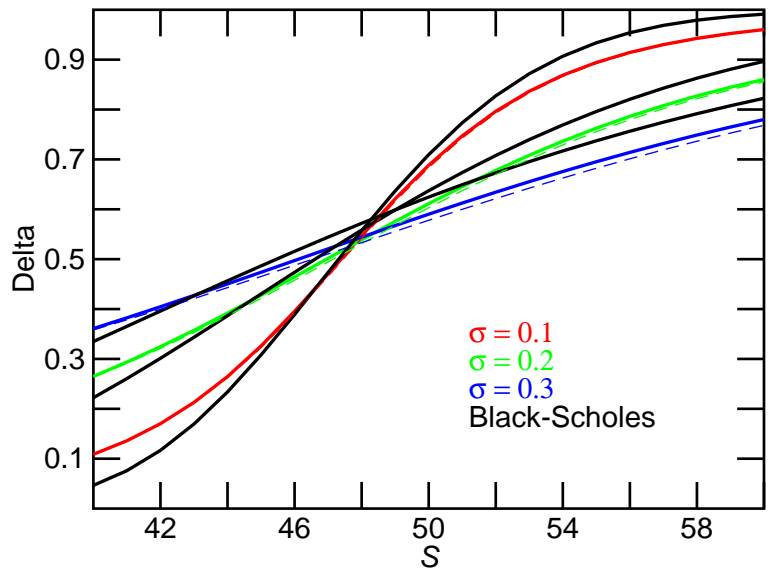}
\caption{Delta as a function of the value of the asset S for different volatilities.}
\end{center}
\end{figure}

\hspace*{1.28cm}An expression for {\it $_{{\rm t}}$\/}$\Gamma$ is given below.  This expression reduces to the Black-Scholes formula for no truncation and $\nu$ = $\infty$. 

\begin{equation}
{\phantom{\,} }_{\it t}\Gamma ={\frac{{\partial\, } ^{2} {\phantom{\,} }_{t}  C_{0} }{ \partial  {\, S_{0} } ^{2} }} ={\frac{K_{T}  \times e^{-r \times  T}  \times f_{r}  \left (\,{\frac{\ln  ({\phantom{\,}}_{t} Z \times  K_{T}  /S_{0} ) - r \times  T\, }{\sigma_{T} }} \right )}{{S_{0} }^{2} \times  \sigma_{T}  \times  (\, p_{c} -p_{p}  )}}
\end{equation}

\hspace*{1.28cm}Figure 8 plots $_{{\rm {\it t\/}}}\Gamma$ for the same parameters as used to create Fig. 5.  The value of $_{{\rm {\it t\/}}}\Gamma$ decreases as the shape parameter decreases.  $_{{\rm {\it t\/}}}\Gamma$ is also a function of {\it p$_{{\rm c}}$\/} and $\sigma_{{\rm {\it T\/}}}$. 
Figure 9 plots $_{{\rm {\it t\/}}}\Gamma$ for $\nu$ = 3{\it , p$_{{\rm p}}$\/} = 0, {\it p$_{{\rm c}}$\/} = 0.999, {\it K$_{{\rm T}}$\/} = \$49.00, {\it r\/}$\times${\it T\/} = 0.03 as a function
of {\it S\/}$_{{\rm 0}}$ for $\sigma_{{\rm {\it T\/}}}$ = 0.1, 0.2, and 0.3 \, The dashed lines gives $_{{\rm {\it c\/}}}\Gamma$, which is $\Gamma$ for a capped asset.  The expression for $_{{\rm {\it c\/}}}\Gamma$ is

\begin{equation}
{\phantom{\,} }_{c}\Gamma ={\frac{{\partial\, } ^{2} {\phantom{\,}}_{c} C_{0} }{ \partial  {\, S_{0} } ^{2} }} ={\frac{K_{T} \times e^{-r \times  T}  \times f_{r}  \left (\,{\frac{\ln  ({\phantom{\,} }_{c}Z \times  K_{T}  /S_{0} ) - r \times  T }{\sigma_{T} }} \right )}{{S_{0} }^{2} \times  \sigma_{T} \, }}
\end{equation}
which is similar to the expression for $_{{\rm {\it t\/}}}\Gamma$.

\begin{figure}
\begin{center}
\label{fig:fig7}
\includegraphics[scale=0.5]{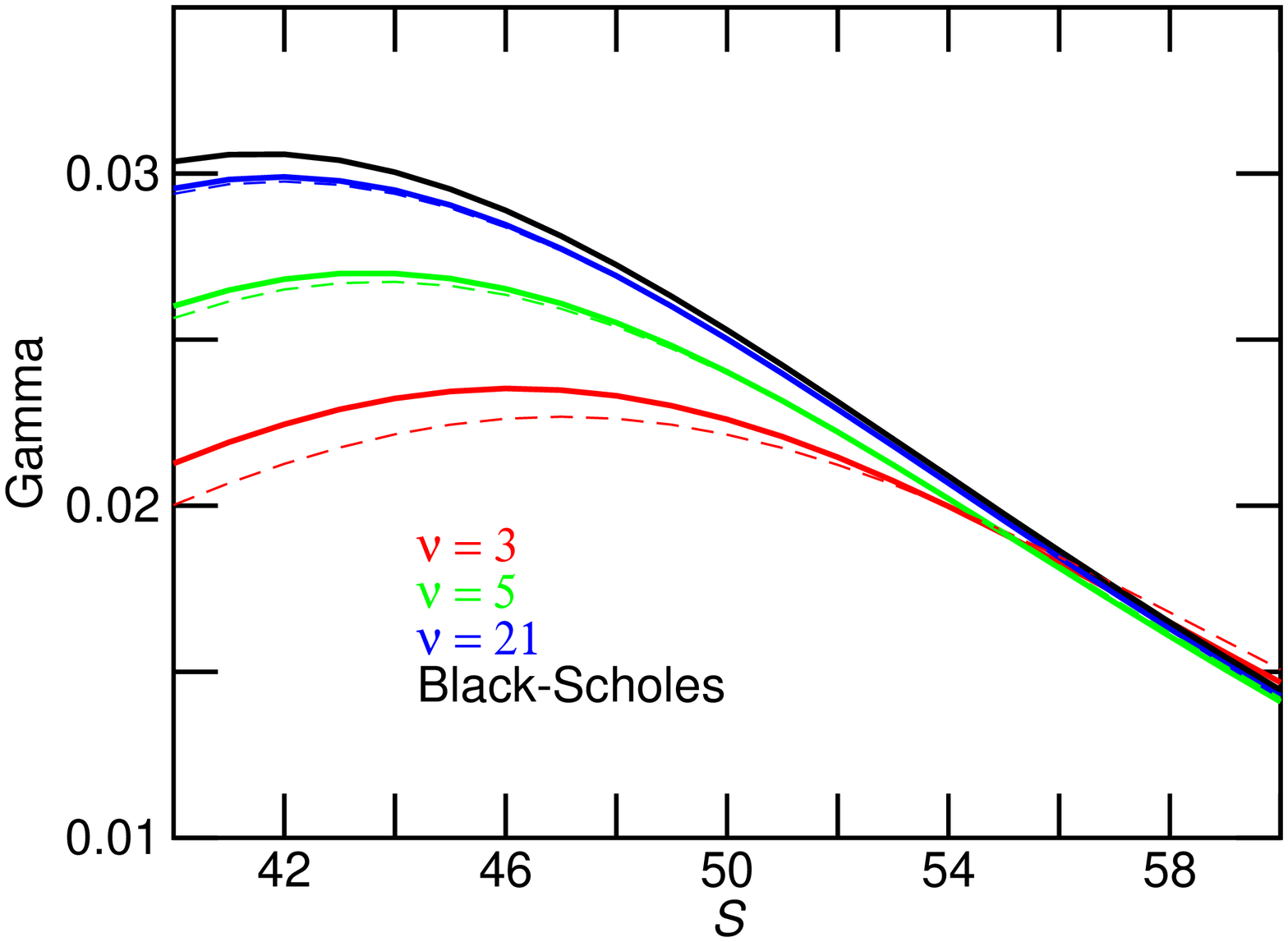}
\caption{Gamma as a function of the value of the asset S for different values of the
number of the degrees of freedom $\nu$.}
\end{center}
\end{figure}

\begin{figure}
\begin{center}
\label{fig:fig8}
\includegraphics[scale=0.5]{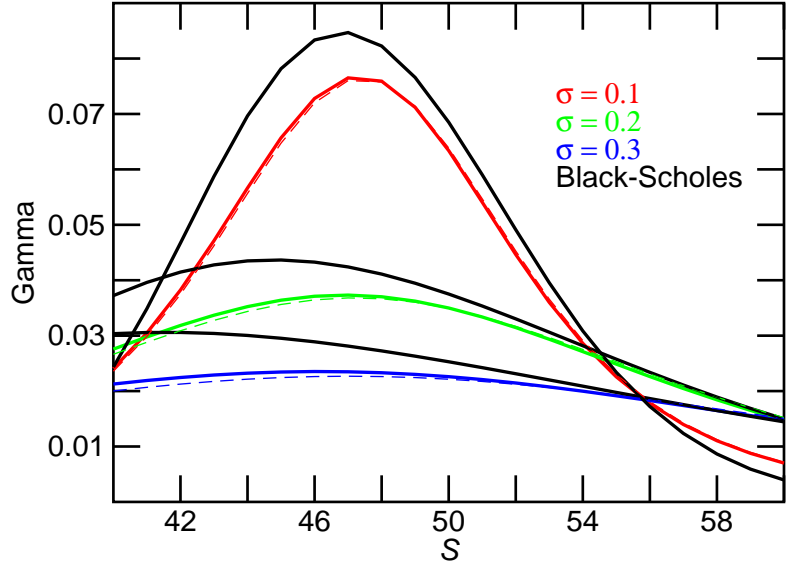}
\caption{Gamma as a function of the value of the asset $S$ for different volatilities.}
\end{center}
\end{figure}

\hspace*{1.28cm}An expression for Vega for a truncated {\it t\/}-distribution, {\it $_{{\rm \it t}}$\/}V, is given below.

\begin{equation}
\begin{array}{r@{}l} {\phantom{\,} }_{t}  V={\frac{\partial  {\phantom{\,} }_{t}C_{0} }{ \partial\, \sigma_{T} }} =&~{\frac{~S_{0} }{{\phantom{\,} }_{t}  Z \times (\, p_{c}  - p_{p} \, )}} \int_{{\frac{\ln  ({\phantom{\,} }_{t}  Z K_{T} /S_{0}  ) - r \times  T}{\sigma_{T} }}}^{x_{c} }  \xi \, e^{( \sigma_{T} \, \xi  )} f_{r} (\xi )\, d \xi  \\
-&~{\frac{S_{0} }{{\phantom{\,} }_{t} Z^{2}  \times ( p_{c} -p_{p} ) }} ~{\frac{ \partial  {\phantom{\,} }_{t} Z}{ \partial\,\, \sigma_{T} }} ~\int_{{\frac{\ln  ({\phantom{\,} }_{t} Z\, K_{T}  /S_{0}  ) - r \times  T}{\sigma_{T} }}}^{x_{c} } e^{(\, \sigma_{T} \, \xi  )} \, f_{r}  (\xi )\, d \xi  \\
{\frac{ \partial  {\phantom{\,} }_{t} Z}{ \partial\,\, \sigma_{T} }} =&~{\frac{~1}{(p_{c} -p_{p} )}} \int_{x_{p} }^{x_{c} } \xi \, e^{(\, \sigma_{T} \, \xi )} \, f_{r} (\xi )\, d \xi \end{array}
\end{equation}
\hspace*{1.28cm}Figure 10 is composed of plots of Vega as a function of {\it S\/}$_{{\rm 0}}$ for the standard parameters used in this work.  The expression for Vega has a term owing to the {\it Z\/}
in the limits of integration.  This term can not be neglected.  Vega increases as the
shape parameter for the {\it t\/}-distribution decreases.  The curve for $\nu$ = 40 lies just
above the curve for the Black-Scholes Vega.  The dashed lines are for Vega for a
capped distribution.  An equation for Vega for a capped asset, {\it $_{{\rm c}}$V\/}, is

\begin{equation}
\begin{array}{r@{}l} {\phantom{\,} }_{c}  V = {\frac{\partial  {\phantom{\,} }_{c} C_{0} }{ \partial\, \sigma_{T} }} =&{\frac{~S_{0} }{{\phantom{\,} }_{c} Z }} \int_{{\frac{\ln \, ({\phantom{\,} }_{c} Z\, K_{T} \, /S_{0} \, ) - r \times  T}{\sigma_{T} }}}^{x_{c} } \, \xi \, e^{(\, \sigma_{T} \, \xi  )} \, f_{r} (\xi )\, d \xi  \\
-&{\frac{~S_{0} }{{\phantom{\,} }_{c} Z^{2} }} ~{\frac{ \partial  {\phantom{\,} }_{c} Z}{ \partial\,\, \sigma_{T} }} ~~\int_{{\frac{\ln \, ({\phantom{\,} }_{c} Z\, K_{T} \, /S_{0} \, ) - r \times  T}{\sigma_{T} }}}^{x_{c} } e^{(\, \sigma_{T} \, \xi \, )} \, f_{r}(\xi )\, d \xi  \\
+&~{\frac{S_{0} \times  e^{\, (\sigma_{T} \, x_{c} )} \times  (1-p_{c} )}{{\phantom{\,} }_{c} Z }} \times  \left ( x_{c} -{\frac{1}{{\phantom{\,} }_{c} Z }}{\frac{ \partial  {\phantom{\,} }_{c} Z}{ \partial\,\, \sigma_{T} }} \right ) \\
{\frac{ \partial  {\phantom{\,} }_{c} Z}{ \partial\,\, \sigma_{T} }} =&~p_{p} \times x_{p} \, e^{(\sigma_{t} \, x_{p} )} +\int_{x_{p} }^{x_{c} } \, \xi \, e^{(\, \sigma_{T} \, \xi  )} \, f_{r} (\xi )\, d \xi + (1-p_{c} )\times x_{c} \, e^{(\sigma_{t} \, x_{c} )}  \end{array}
\end{equation}

\begin{figure}
\begin{center}
\label{fig:fig9}
\includegraphics[scale=0.5]{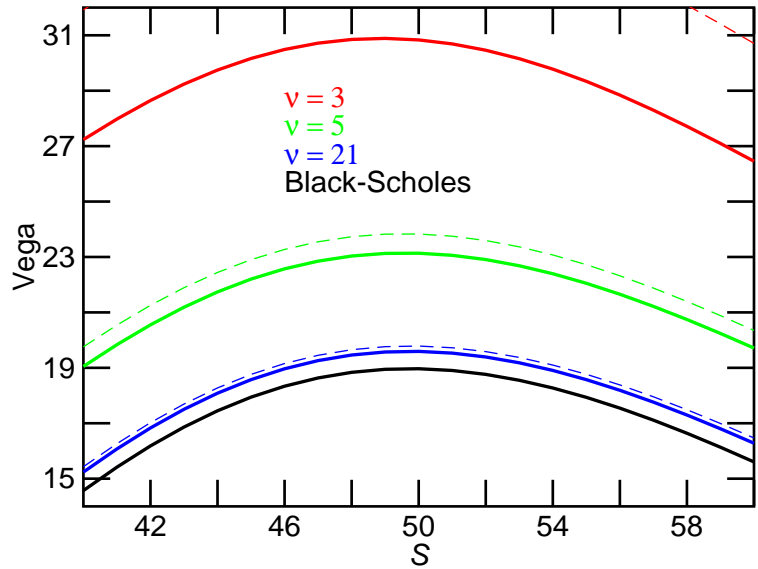}
\caption{Vega as a function of the value of the asset $S$ for different values of the
number of the degrees of freedom $\nu$.}
\end{center}
\end{figure}

\bigskip
\hspace*{1.28cm}The derivatives are messy.  The number of terms and parameters makes it
difficult to determine by looking at the expressions the importance of terms and
parameters.  As a practical matter, it is probably easier to compute the derivatives
numerically as [{\it $_{{\rm \it t}}$C\/}$_{{\rm 0}}$($\zeta_{{\rm 1}}$, $\zeta_{{\rm 2}}$, ...  $\zeta${\it $_{{\rm i}}$\/}+$\epsilon$, $\zeta${\it $_{{\rm i}}$\/}$_{{\rm +1}}$, ...) $-$ $_{{\rm {\it t\/}}}${\it C\/}$_{{\rm 0}}$($\zeta_{{\rm 1}}$, $\zeta_{{\rm 2}}$, ...  $\zeta${\it $_{{\rm i}}$\/}, $\zeta${\it $_{{\rm i}}$\/}$_{{\rm +1}}$, ...) ]/$\epsilon$ for a range
of parameters and plot the values to build an understanding.

\hspace*{1.28cm}Figure 11 is plots of $\Theta$ as calculated numerically as a function of {\it S\/}$_{{\rm 0}}$ for the
standard parameters.  Theta increases as the shape parameter of the Student's {\it t\/}-distribution decreases.  The dashed lines are for a capped distribution.  Theta was
calculated as {\it C\/}$_{{\rm 0}}$({\it p\/}, {\it S\/}, {\it K\/}, {\it r\/}$\times$(366/365), $\sigma${\it $_{{\rm T}}$\/}$\times$(366/365)$^{{\rm 0.5}}$, $\nu$) $-$ {\it C\/}$_{{\rm 0}}$({\it p\/}, {\it S\/}, {\it K\/}, {\it r\/}, $\sigma${\it $_{{\rm T }}$\/}, $\nu$).

\begin{figure}
\begin{center}
\label{fig:fig10}
\includegraphics[scale=0.5]{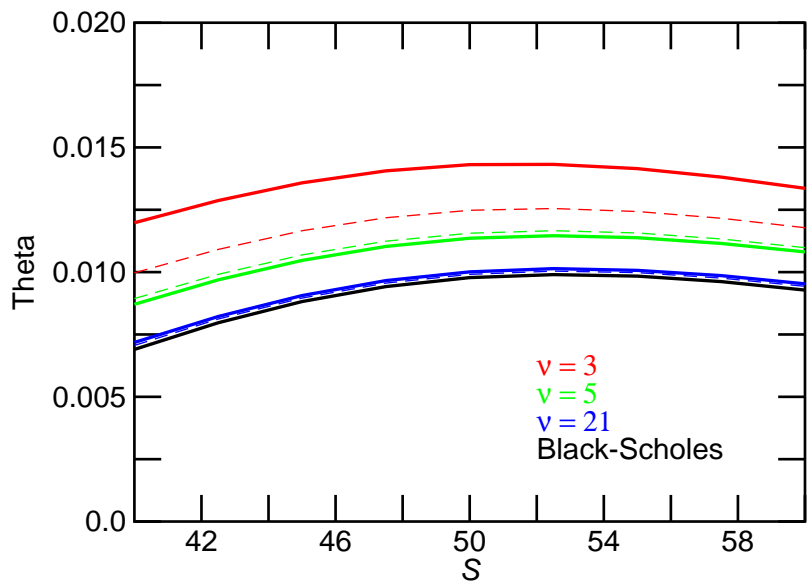}
\caption{Theta as a function of the value of the asset S for different values of the
number of the degrees of freedom $\nu$.}
\end{center}
\end{figure}

\hspace*{1.28cm}Figure 12 is plots of the numerical derivative of the price of a call with
respect to the shape parameter $\nu$ as a function of {\it S\/}$_{{\rm 0}}$ for the standard parameters. 
This derivative is zero for an option priced with the Black-Scholes formula.  Not
surprisingly, the price of a call is sensitive to the shape parameter for small $\nu$.  The
dashed lines are for a capped distribution.

\begin{figure}
\begin{center}
\label{fig:fig11}
\includegraphics[scale=0.5]{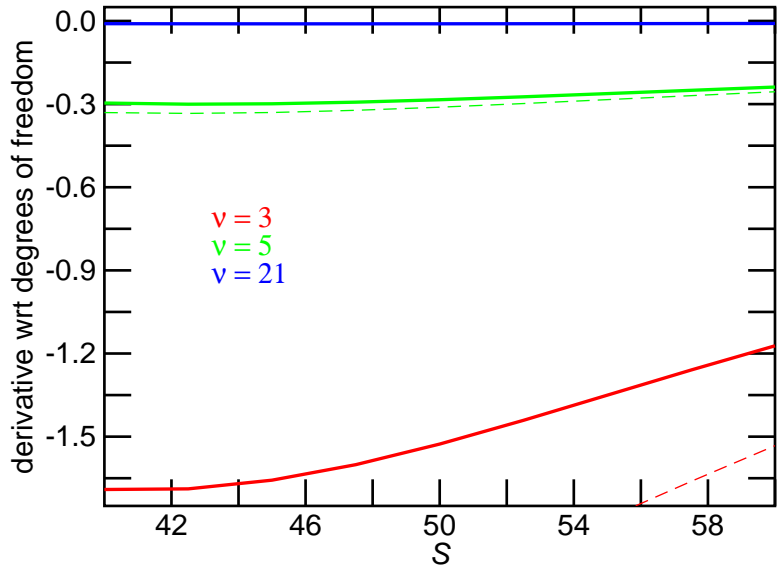}
\caption{Derivative of the price of a call option with respect to the number of
degrees of freedom $\nu$ as a function of the value of the asset $S$ for different values of
the number of the degrees of freedom.}
\end{center}
\end{figure}

\hspace*{1.28cm}Figure 13 is plots of the logarithm of the derivative of the price of a call with
respect to the truncation {\it p = p$_{{\rm c}}$\/} as a function of {\it S\/}$_{{\rm 0}}$ for the standard parameters.  The
dashed lines are for a capped distribution.  Note the large value of the magnitude of
the derivative for {\it p\/} approaching unity.  Note also that the maximum value for $\epsilon$ is
$<<$ (1 $-$ {\it p\/}) and that the change in the price of a call for a change in {\it p\/} of $\epsilon$ is the derivative$\times\epsilon$.  

\begin{figure}
\begin{center}
\label{fig:fig12}
\includegraphics[scale=0.5]{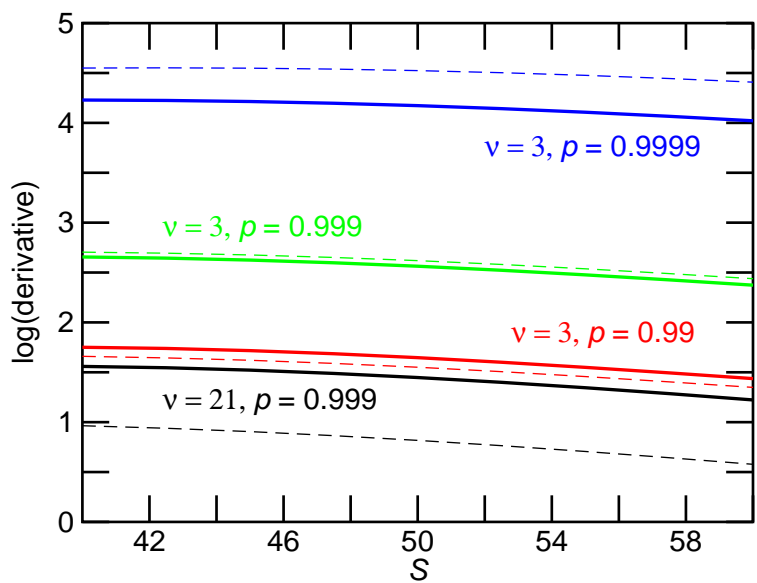}
\caption{Plots of the logarithm of the derivative of the value of a call option with
respect to the level of truncation $p$ as a function of the value of the asset $S$ for
different values for the level of truncation.}
\end{center}
\end{figure}

\hspace*{1.28cm}The integrals that are required to price a call option contain terms with
exp($\sigma_{{\rm {\it T\/}}}$ {\it x$_{{\rm c}}$\/}).  As {\it p$_{{\rm c}}$\/} approaches unity, {\it x$_{{\rm c}}$\/} approaches infinity, and the integrals
required to price a call option assuming the returns follow a Student's {\it t\/}-distribution
approach infinity.  The case of {\it x$_{{\rm c}}$\/} approaching infinity is unphysical as this suggests
that the value of the asset exp($\sigma_{{\rm {\it T\/}}}$~$\times~\infty$) must also approach infinity.  As {\it p$_{{\rm c}}$\/}
approaches unity, the price of the option increases to account for the values of the
asset that are being included.

\bigskip

\section{Conclusion}

\hspace*{1.28cm}It is known that a Student's {\it t\/}-distribution fits the returns of stocks better than a normal pdf.  A Student's {\it t\/}-distribution has ``fat tails'', and the shape parameter $\nu$
and scale parameter $\sigma_{{\rm {\it T\/}}}$ of the {\it t\/}-distribution can be selected to match the fat tails of the returns. 

\hspace*{1.28cm}Prices for European options can be determined for assets that follow a log
Student's {\it t\/}-distribution if the value of the asset is capped or if the {\it t\/}-distribution is
truncated.  Capping or truncating the distribution keeps the integrals that are
required to price an option finite.  We call the formulae that use a Student's {\it t\/}-distribution to calculate the values of options Gosset formulae, in honour of W.S.
Gosset.  Gosset used the pseudonym Student to publish his work.

\hspace*{1.28cm}In the limit of large $\nu$ the Gosset formulae converge to the Black-Scholes
formula for the values of European options.  The Gosset formula can be viewed as
an extension of the Black-Scholes formula to allow for a volatility $\sigma_{{\rm {\it T\/}}}$ that has
uncertainty or is a random variable.  The Black-Scholes formula is based on the
assumption that the volatility is known and constant.  The returns must be normally
distributed and the volatility must follow a chi or inverse chi distribution for the
Student's {\it t\/}-distribution to hold strictly.  The historical returns from the DJIA and
S\&P 500 appear to follow these distributions as the Student's {\it t\/}-distribution fits
well the returns on these indexes.

\hspace*{1.28cm}It is necessary to know the shape parameter $\nu$ to use the Gosset formulae.  

\hspace*{1.28cm}The maximum value of $\nu$ $<$ {\it N\/} where {\it N\/} is the number of independent returns
that are used to calculate the volatility.  The shape parameter, in general, will be
less than the maximum value for large {\it N\/} (say {\it N\/} $>$ 22).  The volatility is not
constant in time, and for large {\it N\/}, the uncertainty of the volatility broadens the pdf
that describes the returns and hence reduces $\nu$.  Also, for longer time to expiration
{\it T\/}, the volatility at time {\it T\/} is known with less accuracy, and a reduced $\nu$ is necessary
to capture the uncertainty in the future value of the volatility.  Fits to the returns
from the DJIA (Oct. 1928 to Feb. 2009) and the S\&P 500 (Jan. 1950 to Feb. 2009)
give $\nu$ of order 3.  Thus it appears that typically $\nu$ should be in the range of $\sim$3 to
{\it N\/}$-$1.  A method to estimate an appropriate value of $\nu$ by comparison of the
standard deviations of the returns as the minimum and maximum values are
dropped is given. 

\hspace*{1.28cm}Some of the greeks for the Gosset and Black-Scholes formulae for pricing
European options are given and compared.  The differences between the two
approaches are the largest for small $\nu$.  This is not surprising, given that the
difference between the two underlying distributions for the returns are the greatest
for small $\nu$.  

\hspace*{1.28cm}New greeks are required to characterize the price sensitivity to parameter
change for the Gosset approach.  The Gosset formula requires a cap or truncation
and a shape parameter $\nu$.  Changes in the value of a call option for changes in $\nu$ and
the truncation or cap level are presented.

\newpage

\appendix
\section{}  

\begin{flushleft}
A Maple script to price a call option for 100 different values of the asset price is
presented in this Appendix.  The script was used to find the execution time that is
reported in Sec. 2, Implementation.  The intrinsic function to price a call with
the Black-Scholes formula is included as a comment in the loop.
\end{flushleft}

\begin{figure}[h!]
\begin{center}
\label{fig:script}
\includegraphics[scale=0.7]{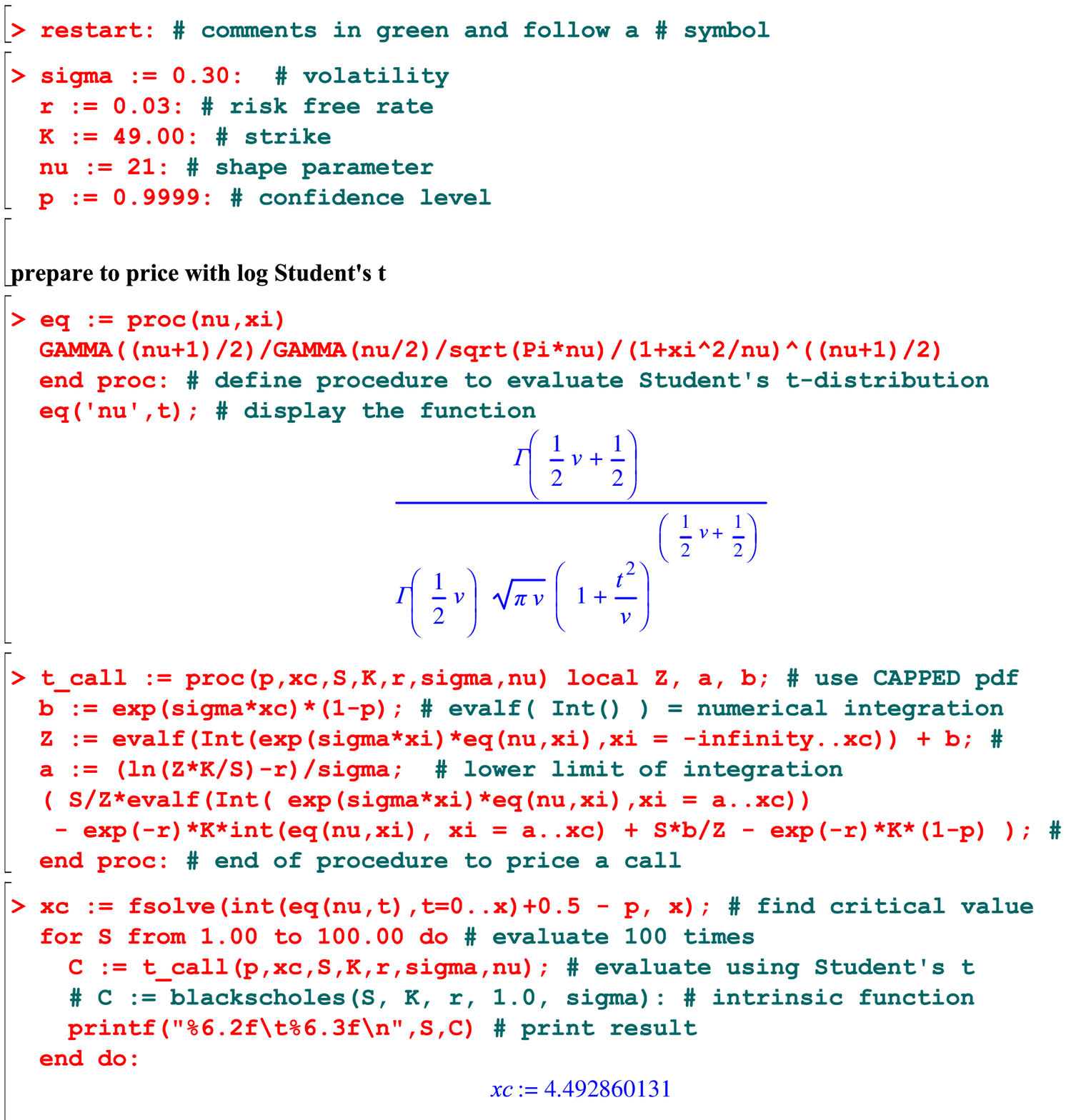}
\caption{Maple script for Appendix A.}
\end{center}
\end{figure}

\newpage
\section{} 
\begin{flushleft}
{\bf \hspace*{1.28cm}}The equations to price an option by capping the value of an asset or truncating the underlying pdf are presented in a general form.  The approach follows the work of Cassidy, Hamp, and Ouyed \citep{cas}.

\end{flushleft}

\subsection{Pricing the Option}

\hspace*{1.28cm}Let {\it S$_{{\rm\it t}}$\/} be the price of a stock at time {\it t\/}, {\it t\/} $>$ 0.  Let {\it K$_{{\rm\it T}}$\/} be the strike price at time {\it T,\/} where {\it T\/} is the time when the option expires.

\hspace*{1.28cm}Let {\it S$_{{\rm\it t}}$\/} = {\it A$_{{\rm\it t }}$\/}exp($\sigma_{{\rm {\it t\,\/}}}\xi$) be the value of the stock where $\xi$ is a random variable and let the probability density function for the return be {\it f$_{{\rm\it r}}$\/}($\xi$).  If $\xi$ is a Student's {\it t\/} variate, then

\begin{equation}
{f}_{r} (\xi )\, d \xi = {\frac{\Gamma \, ({{ (\nu +1)}/ {2}   )}}{\Gamma (\nu /2)\, \sqrt{\pi \nu } }} \times{\frac{d\xi }{{\left ( 1+{\frac{\xi^{2} }{\nu }} \right )}^{{\frac{v+1}{2}}} }}
\end{equation}
\hspace*{1.28cm}The value of a European call option, calculated at the time of expiration {\it T\/}, is
{\it C$_{{\rm\it T}}$\/} = E\{({\it S$_{{\rm\it T}}$\/}~$-$ {\it K$_{{\rm\it T}}$\/})$^{{\rm +}}$\}, which is the expectation of the maximum value of \{{\it S$_{{\rm\it T}}$\/} $-$ {\it K$_{{\rm\it T }}$\/}, 0\}.  The value of a European put option is {\it P$_{{\rm\it T}}$\/} = E\{({\it K$_{{\rm\it T}}$\/}~$-$ {\it S$_{{\rm\it T}}$\/})$^{{\rm +}}$\}.  These expressions for the values of the options follow from the arbitrage theorem \citep{ros}.  The values of the options at time {\it t\/} = 0 are obtained from the expected time value of money.  If {\it r\/}({\it t\/}) is the risk free rate, then {\it C\/}$_{{\rm 0}}$ = E\{{\it C$_{{\rm\it T}}$\/} $\times$ exp($-\int${\it r\/}({\it t\/})d{\it t\/})\} = {\it C$_{{\rm\it T}}$\/} $\times$ exp($-${\it r\/}$\times${\it T\/}) when the risk free rate is assumed to be time independent, with a similar equation for {\it P$_{{\rm\it T}}$\/}.  

\hspace*{1.28cm}The average value of a stock at time {\it t\/} = 0 when one wishes to price an option
is {\it S\/}$_{{\rm 0}}$.  The average value of the stock at some time later is E\{{\it S$_{{\rm\it t}}$\/}\} = E\{{\it A$_{{\rm\it t }}$\/}exp($\sigma_{{\rm {\it t\,\/}}}\xi$)\}
where the drift is contained in {\it A$_{{\rm\it t}}$\/}.  For a martingale (a fair wager), E\{{\it S$_{{\rm\it t}}$\/}\} = {\it S\/}$_{{\rm 0}}\times$exp({\it r$\times$t\/}).  One may use the Doob decomposition to offset the drift and require E\{{\it A$_{{\rm\it t }}$\/}exp($\sigma_{{\rm {\it t\,\/}}}\xi$)\} = E\{{\it A$_{{\rm\it t}}^\prime$\/}exp($\sigma_{{\rm {\it t \/}}}$($\xi-\xi_{{\rm o}}$))\} = {\it S\/}$_{{\rm 0}}\times$exp({\it r$\times$t\/}).  This is equivalent to requiring that the pdf for {\it S$_{{\rm\it t}}$\/} should be centred about $\xi_{{\rm o}}$, the drift owing to the risk premium.  $\xi_{{\rm o}}$ is not a random variable.  The exp({\it r$\times$t\/}) term takes into account the time value of money and allows for comparison of values at two different points in time. 

\hspace*{1.28cm}The difficulty with pricing an option for which the return follows a Student's
{\it t\/}-distribution is that E\{{\it S$_{{\rm\it t}}$\/}\} is infinite.  The exp($\sigma_{{\rm {\it t\, \/}}}\xi$) term in the expectation dominates over the $\xi^{{\rm -}{}{\rm \nu}{}}$ behaviour of the pdf for large $\xi$.  Two methods for pricing an option are presented below.  One method is to place a cap on the value of the asset. The other method is to truncate the pdf.  Truncation is physical because it is not possible for the price of an asset to approach infinity.  The equations in this
appendix are written in general terms.  The approach to price an option by capping
or truncating is not restricted to a Student's {\it t\/}-distribution.  The approach can be
used with other probability density functions.  In this appendix, provision is made
for a floor {\it x$_{{\rm \it p}}$\/} or for a truncation at the floor.

\hspace*{1.28cm}Let {\it p$_{{\rm\it p}}$\/} = P\{$\xi$ $\leq$ {\it x$_{{\rm\it p}}$\/}\} and {\it p$_{{\rm\it c}}$\/} = P\{$\xi$ $\leq$ {\it x$_{{\rm\it c}}$\/}\} where {\it x$_{{\rm\it c}}$\/} and {\it x$_{{\rm\it p}}$\/} are critical values.  In
terms of the pdf, {\it p$_{{\rm\it p}}$\/} and {\it p$_{{\rm\it c}}$\/} are equal to (with {\it i\/} = {\it p\/} or {\it c\/})

\begin{equation}
p_{i} =\int_{-\infty }^{x_{i} }  {f }_{r}  (\xi )\, d\xi
\end{equation}
\hspace*{1.28cm}For {\it p$_{{\rm\it p}}$\/} = 0, {\it p$_{{\rm\it c}}$\/} = 1, and {\it f$_{{\rm\it r}}$\/}($\xi$) a standard normal variate, the equations for the
prices of European options presented reduce to the Black-Scholes formula.  We call
the formulae for prices of options that have been obtained with {\it f$_{{\rm\it r}}$\/}($\xi$) a Student's {\it t\/}-distribution Gosset formulae, in honour of W. S. Gosset \citep{zab}.

\bigskip
\subsection{Cap and Floor}

 \bigskip
\hspace*{1.28cm}Assume a floor {\it S$_{{\rm\it T}}$\/} = {\it x$_{{\rm\it p}}$\/} for {\it S$_{{\rm\it T}}$\/} $\leq$ {\it x$_{{\rm\it p}}$\/} and a cap {\it S$_{{\rm\it T}}$\/} = {\it x$_{{\rm\it c}}$\/} for {\it S$_{{\rm\it T}}$\/} $\geq$ {\it x$_{{\rm\it c}}$\/} and let 

\begin{equation}
{\phantom{a} }_{c} Z=p_{p} \times e^{(\sigma_{T} x_{p} )} +\int_{x_{p} }^{x_{c} } e^{(\sigma_{T} \, \xi }  \times {f}_{r} (\xi ) \times d \xi +(1-p_{c} ) \times e^{(\sigma_{T} \, x_{c} )} 
\end{equation}
\hspace*{1.28cm}A pre-subscript of {\it c\/} or {\it t\/} indicates that a quantity has been calculated with a
cap or a truncation.  One could write {\it Z\/} as $_{{\rm\it f{\it c\/}}}${\it Z\/} to indicate a floor as well as a cap. 
However, the extra symbol is not necessary here and is omitted.

\hspace*{1.28cm}For a fair wager, the expected value of {\it S$_{{\rm\it T}}$\/} is found as

\begin{equation}
{E \{S_{T} \} = S_{0} \times e^{\, (r \times T )} = {}_{c}A_{T} \times _{c}Z}
\end{equation}
and the unknown {\it $_{{\rm\it c }}$A$_{{\rm\it T}}$\/} can be found

\begin{equation}
{\phantom{i} }_{c} A_{T} = {\frac{S_{0} \times\, e^{(r \times  T )} }{ {\phantom{~} }_{c} Z}}
\end{equation}
\hspace*{1.28cm}The price of a European call option for a capped asset value is

\begin{equation}
\begin{array}{r@{}l} {\phantom{a} }_{c} C_{T} =&~ \int_{{\frac{\ln (K_{T} /{\phantom{i} }_{c} A_{T} )}{\sigma_{T} }} }^{x_{c} } \left ( {\phantom{i} }_{c} A_{T} \, e^{\, (\sigma_{T} \, \xi  )} -K_{T} \right )  \times {f }_{r} (\xi  ) \times d \xi  \\
+&~{(1-p_{c} ) \times ( {\phantom{i} }_{c} A_{T} \, e^{\, (\sigma_{T} x_{c} )} -K_{T} \, )}~ \end{array}
\end{equation}
assuming that ln({\it K$_{{\rm\it T }}$\/}/{\it $_{{\rm\it c }}$A$_{{\rm\it T}}$\/})/$\sigma_{{\rm\it {\it T \/}}}$ $>$ {\it x$_{{\rm\it p }}$\/}.

\hspace*{1.28cm}The price of a European put for a capped asset value is

\begin{equation}
\begin{array}{r@{}l} {\phantom{a} }_{c} P_{T} =&~ p_{p} \times (K_{T} - {\phantom{i}}_{c} A_{T} \, e^{\, (\sigma_{T} \, x_{p} )}  \, )  \\
+&~ \int_{x_{p}}^{\frac{\ln (K_{T}/{\phantom{i}}_{c}A_{T} )}{\sigma_{T}} } \left ( K_{T} - {\phantom{i} }_{c}A_{T} \, e^{\, (\sigma_{T} \, \xi \, )}  \right )  \times  {f}_{r}(\xi) \times d\xi ~ \end{array}
\end{equation}

\hspace*{1.28cm}In the limit as {\it p$_{{\rm\it p}}$\/} tends to zero, the floor disappears.  Note that a cap or
truncation is necessary to price an option with a Student's {\it t\/}-distribution.  The cap
or truncation keeps the integrals, which are needed to price the options, finite.  The
floor is not necessary but may make for realistic pricing.  The price of an option is
quantized by the currency.  For values less than the smallest unit of the currency, it
may be appropriate to set the price equal to zero, i.e., to use a floor.

\hspace*{1.28cm}Put-call parity holds for truncated or capped options.  The equation for the
price of a put is presented for completeness.

\bigskip
\subsection{Truncation}

\hspace*{1.28cm}Assume that {\it S$_{{\rm\it T}}$\/} only takes values {\it x$_{{\rm\it p}}$\/} $\leq$ {\it S$_{{\rm\it T}}$\/} $\leq$ {\it x$_{{\rm\it c}}$\/} and thus {\it f$_{{\rm\it r}}$\/}($\xi$) = 0 if $\xi$ $<$ {\it x$_{{\rm\it p}}$\/} or $\xi$ $>$
{\it x$_{{\rm\it c }}$\/}.  Let

\begin{equation}
{\phantom{a} }_{t} Z=\int_{x_{p} }^{x_{c} } \,{\frac{~e^{\, (\sigma_{T} \, \xi  )} \times {f}_{r} (\xi ) \times d \xi ~}{p_{c} - p_{p} }}
\end{equation}

 \bigskip
The denominator follows from the definition of conditional probability and is
required to maintain the normalization of the pdf after the truncation:
P\{{\it x\/}~$<$~$\xi$~$\leq$~{\it x\/}+{\it dx\/}~$|$~{\it x$_{{\rm\it p}}$\/}~$\leq$~$\xi$~$\leq$~{\it x$_{{\rm\it c}}$\/}\} = P\{{\it x\/}~$<$~$\xi$~$\leq$~{\it x\/}+{\it dx\/}~$\cap$~{\it x$_{{\rm\it p}}$\/} $\leq$ $\xi$~$\leq$~{\it x$_{{\rm\it c}}$\/}\}/P\{{\it x$_{{\rm\it p}}$\/} $\leq$ $\xi$ $\leq$ {\it x$_{{\rm\it c }}$\/}\} =
{\it f$_{{\rm\it r}}$\/}($\xi$)d{\it x\/}/({\it p$_{{\rm\it c}}$\/} $-$ {\it p$_{{\rm\it p}}$\/}) if {\it x$_{{\rm\it p}}$\/} $\leq$ $\xi$ $\leq$ {\it x$_{{\rm\it c}}$\/} and zero otherwise.

\hspace*{1.28cm}The average value of the stock at {\it T\/} is E\{{\it S$_{{\rm\it T}}$\/}\} = {\it S\/}$_{{\rm\it 0}}\times$exp({\it r$\times$t\/}) {\it = $_{{\rm\it t }}$A$_{{\rm\it T}}$\/} $\times$ $_{{\rm\it {\it t \/}}}$Z or

\begin{equation}
_{t} A_{T} = {\frac{S_{0} \times e^{\, (r \times T )}} { _{t} Z }}
\end{equation}
\hspace*{1.28cm}The price of a European call option for a truncated pdf is

\begin{equation}
_{t} C_{T} = \int_{{\frac{\ln (K_{T} /_{t} A_{T} )}{\sigma_{T} }} }^{x_{c} } {\frac{\left ( _{t} A_{T} \, e^{\, (\sigma_{T} \, \xi \, )} - K_{T} \, \right ) \times  {f}_{r} (\xi ) \times d \xi }{p_{c} - p_{p} }}
\end{equation}
assuming that ln({\it K$_{{\rm\it T }}$\/}/$_{{\rm {\it t \/}}}${\it A$_{{\rm\it T}}$\/})/$\sigma_{{\rm {\it T \/}}}>$ {\it x$_{{\rm\it p }}$\/}.

\hspace*{1.28cm}The price of a European put for a truncated pdf is 

\begin{equation}
{ _{t} P_{T} =\int_{{x_{p} }}^{{\frac{\ln (K_{T} /_{t} A_{T} )}{\sigma_{T} }} } {\frac{\left ( K_{T} -\, _{t} A_{T} \, e^{\, (\sigma_{T}\, \xi \, )}  \, \right ) \times  {f}_{r} (\xi ) \times  {\rm d}\xi }{ p_{c} - p_{p} }}  }
\end{equation}
\hspace*{1.28cm}Note that $_{{\rm {\it t \/}}}${\it A$_{{\rm\it T}}$\/} is known in terms of {\it S\/}$_{{\rm 0}}$, {\it r\/}, and {\it $_{{\rm\it t}}$Z\/}({\it p$_{{\rm\it p}}$\/}({\it x$_{{\rm\it p}}$\/}), {\it p$_{{\rm\it c}}$\/}({\it x$_{{\rm\it c}}$\/})) and thus the
option prices can be determined.  Note also that {\it $_{{\rm\it t}}$Z\/} $\neq$ {\it $_{{\rm\it c}}$Z\/}.

\bigskip

\section*{\bf Acknowledgements}

\hspace*{1.28cm}This work was funded in part by the Natural Sciences and Engineering
Research Council of Canada.

\bigskip

\bibliographystyle{plainnat} 

\label{lastpage}
\end{document}